\documentclass[aps,prb,showpacs,reprint]{revtex4-1}

\usepackage{graphicx}
\usepackage{amsmath}
\usepackage{bbm}
\usepackage{bm}

\begin{document}
\title
{Electrically controlled spin-transistor operation in helical magnetic field}
\author{P. W\'ojcik}
\email[Electronic address: ]{pawel.wojcik@fis.agh.edu.pl}
\author{J. Adamowski}
\affiliation{AGH University of Science and Technology, Faculty of
Physics and Applied Computer Science, al. Mickiewicza 30,
Krak\'ow, Poland}

\begin{abstract}
A proposal of electrically controlled spin transistor in helical magnetic field is
presented. In the proposed device, the transistor action is driven by the Landau-Zener
transitions that lead to a backscattering of spin polarized electrons and switching the transistor
into the high-resistance state (off state). The on/off state of the transistor can be controlled
by the all-electric means using Rashba spin-orbit coupling that can be tuned by the voltages
applied to the side electrodes.
\end{abstract}

\pacs{}

\maketitle
An application of electron spin to conventional electronic devices is the
basic concept lying in the heart of spintronics.\cite{Szumniak2012,Fabian2007} The special role
among the spintronic devices is played by the spin field effect transistor (spin-FET), the concept
of which was proposed by Datta and Das.\cite{Datta1990} According to this ideas\cite{Datta1990} the
current of spin polarized electrons is injected from the ferromagnetic source into the
two-dimensional electron gas (2DEG) and is electrically detected in the ferromagnetic drain. The
resistance of the transistor channel depends on the electron spin, which is controlled by the Rashba
spin-orbit interaction (SOI)~\cite{Rashba1984} generated by the voltage applied to the gate attached
close to the channel.
Although the coherent manipulation of electron spin in semiconductor have been demonstrated by many
experimental groups,\cite{Koo2009,Crooker2005,Appelbaum2007} the realization of the functional
spin-FET still remains an experimental challenge of spintronics. This is due to the
fundamental physical obstacles such as the low spin-injection efficiency from ferromagnet into
semiconductor (the resistance mismatch),\cite{Schmidt2000} spin relaxation,\cite{Fabian2007}
and the spread of the spin precession angles. All these effects lead to the low electrical signal in
the up-to date realized spin-FET.\cite{Koo2009,Wunderlich2010,Wojcik2015} 

The current research on the spin-FET is mainly focused on two directions. First, the researchers
try to overcome the physical obstacles by using the spin filters based on
semiconductors.\cite{Wojcik2013, Wojcik2015} In the very recent experiment,\cite{Chuang2015} the
quantum point contacts have been used as the spin injectors and detectors and about $100,000$ times
greater conductance oscillations have been observed as compared to the
conventional spin-FET.\cite{Koo2009}
The second direction involves the proposals of the new spin transistor design, different from that
proposed in Ref.~\onlinecite{Datta1990} and free of its physical restrictions.\cite{Schliemann2003,
Kunihashi2012}

Such an alternative spin-transistor design has been recently described by {\v Z}uti{\'c} and
Lee~\cite{Zutic2012} and experimentally demonstrated by Betthausen et al.
in Ref.~\onlinecite{Betthausen2012}. In this approach, the spin of electron flowing through the
nanostructure is subjected to the combined homogeneous and helical magnetic fields. The latter is
generated by ferromagnetic stripes located above the conduction channel. The transistor action is
driven by the diabatic Landau-Zener transitions~\cite{Landau,Zener} induced by the appropriate
tuning of the homogeneous magnetic field. For the suitably chosen conditions
the spin polarized electrons are scattered back, which gives raise to the increase of the
resistance. As shown in Refs.~\onlinecite{Saarikoski2014, Betthausen2012, Wojcik2015_SST}, by
keeping the transport in the adiabatic regime, the proposed design is robust against the scattering
on defects, while in the non-adiabatic regime, the additional conductance dips emerge which result
from the resonant Landau-Zener transitions. Although this spin-FET\cite{Betthausen2012} seems to be
free of the above-mentioned physical obstacles, it requires the application and control of
the external homogeneous magnetic field, which is difficult to be applied in the integrated circuit.

Motivated by this experiment,\cite{Betthausen2012} in the present paper, we propose the spin
transistor design based on the helical magnetic field, in which the spin transistor action is
generated by all-electric means without the external magnetic field. For this purpose
we use the Rashba spin-orbit interaction induced by the lateral electric field between electrodes
located on either side of the channel. 

\begin{figure}[ht]
\begin{center}
\includegraphics[scale=0.35, angle=0]{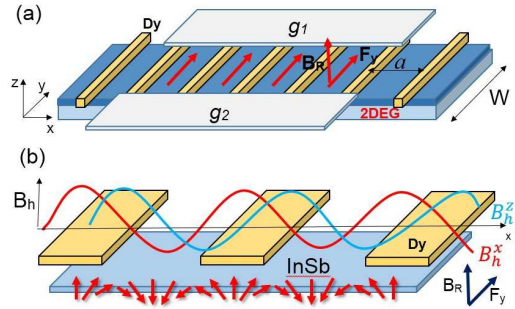}
\caption{(Color online)
(a) Schematic of the proposed spin transistor. 2DEG in InSb layer is subjected to the helical
magnetic field $\mathbf{B}_h$ generated by the ferromagnetic stripes (Dy) located above the
conduction channel, the Rashba SOI is controlled by the side gates $g_1$ and $g_2$.
(b) Profile of the helical magnetic field and spin distribution in 2DEG.
The effective Rashba magnetic field $B_R$ is directed along the $z$ axis. $F_y$ is the electric
field generated by the gates.}
\label{fig1}
\end{center}
\end{figure}

We consider the two-dimensional waveguide (nanowire) in the $x-y$ plane made of InSb,
for which the Rashba spin-orbit coupling is reported  to be strong.\cite{Nadj2012}  The
proposed transistor design (Fig.~\ref{fig1}) consists of the two building blocks: (i) ferromagnetic
stripes (fabricated from dysprosium, Dy) located periodically above the conduction channel, which
generate the helical magnetic field in the
2DEG (this block has been realized in Ref.~\onlinecite{Betthausen2012}) and (ii) two side
gate electrodes ($g_1$ and $g_2$), which generate the lateral electric field $\mathbf{F}=(0,F_y,0)$.
The lateral electric field $\mathbf{F}$, controlled by the voltages applied to the side gates,
induces the Rashba spin-orbit interaction with the effective magnetic field $B_R$ directed along
the $z$-axis. The possibility of using the SOI induced by the lateral
electric field in the QPC has been recently reported in many
experiments~\cite{Chuang2015,Debray2009,Das2011}. The change in
the SOI coupling constant $\alpha$ between the two electrodes has been obtained to be $4-50$~meVnm.

The Hamiltonian of the electron is given by
\begin{equation}
\hat{H}=\frac{\hbar ^2 \mathbf{k}^2}{2m^*} \mathbf{1} + \frac{1}{2}g_{e\!f\!f}\mu_B
\mathbf{B_h}(\mathbf{r}) \cdot \bm{\sigma} + \alpha \sigma _z k_x\;,
\label{ham}
\end{equation}
where $m^*$ is the conduction-band electron mass, $\hbar \mathbf{k}=-i\hbar\nabla$ is the momentum
operator, $\mathbf{1}$ is the $2 \times 2$ identity matrix, $\bm{\sigma}=(\sigma _x,\sigma _y,\sigma
_z)$ is the vector of Pauli matrices, $g_{e\!f\!f}$ is the effective $g$-factor, $\mu_B$ is the Bohr
magneton, $\alpha$ is the Rashba spin-orbit coupling constant and $\mathbf{B_h}(\mathbf{r})$ is the
helical magnetic field, which has the form  
\begin{equation}
\mathbf{B}_h(\mathbf{r})=B_h \left ( \sin {\frac{2 \pi (x-x_0)}{a}}, 0, \cos
{\frac{2 \pi (x-x_0)}{a}} \right ) \;,
\end{equation}
where $B_h$ is the amplitude of the helical field, $a$ is the period of the magnetic field
modulation determined by the distance between ferromagnetic stripes and $x_0$ is the
localization of one of the ferromagnetic stripes. For simplicity, we study the waveguide with $L$ 
length equal to one period of the helical magnetic field modulation $L=a$ and $x_0=L/2$. In the
present paper, $L=1\mu$m.

\begin{figure}[ht]
\begin{center}
\includegraphics[scale=0.3, angle=0]{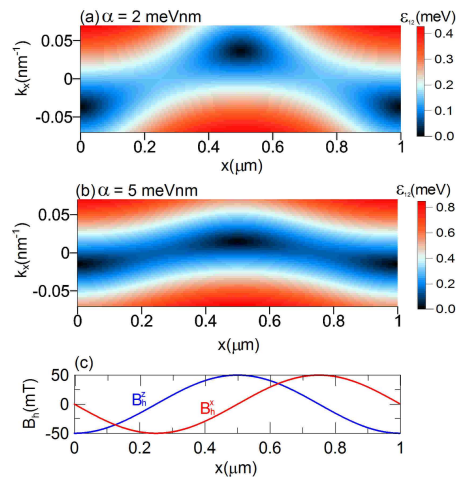}
\caption{(Color online) Energy separation $\varepsilon _{12}$ as a
function position $x$ in the waveguide and wave vector $k_x$ for the Rashba SOI
strength (a) $\alpha=2$meVnm and (b)  $\alpha=5$meVnm. (c) Spatial changes of the helical
magnetic field components $B_h^x$ and $B_h^z$.
}
\label{fig2}
\end{center}
\end{figure}

\begin{figure}[ht]
\begin{center}
\includegraphics[scale=0.3, angle=0]{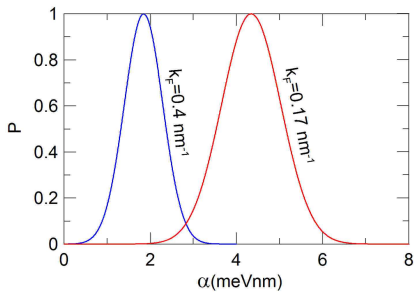}
\caption{(Color online) Landau-Zener transition probability $P$ as a function of Rashba
spin-orbit coupling $\alpha$ for two chosen values of Fermi wave vector $k_F$. 
}
\label{fig3}
\end{center}
\end{figure}

The behavior of the electron spin in the 2DEG is determined by the superposition of the helical
magnetic field $\mathbf{B}_h$ generated by the ferromagnetic stripes and the Rashba effective
magnetic field $\mathbf{B}_R$ induced by the lateral electric field.
The effective spin Zeeman energy for the parallel ($+$) and antiparallel ($-$) spin orientation
(with respect to the effective magnetic field) depends on the electron position and is given by 
\begin{equation}
 E^{\pm}(x)=\pm \alpha k_x \sqrt{1+\gamma ^2 + 2 \gamma \cos \left ( \frac{2\pi (x-x_0)}{a} \right )
} \;,
\end{equation}
where $\gamma=\frac{1}{2}g_{eff} \mu_B B_h/ \alpha k_x$.  Parameter $\gamma$ depends on the Rashba
spin-orbit coupling $\alpha$, which can be tuned by changing the voltages applied to the side
electrodes.\\
For the suitably chosen conditions, in this two level system the diabatic Landau-Zener transitions
can take place with the probability~\cite{Zener, Landau}
\begin{equation}
 P=\exp \left ( - \frac{2 \pi}{\hbar ^2} \frac {\varepsilon _{12}}{ \beta} \right ),
 \label{eq:PLZ}
\end{equation}
where $\varepsilon _{12}=(E^+-E^-)/2$. Parameter $\beta=\frac{1}{\hbar} \frac{d}{dt} \left ( E^+ -
E^- \right )$ determines how fast these eigenenergies approach each other in the electron rest
frame. For our system, $\beta$ takes on the form 
\begin{equation}
\beta =\frac{2\pi g_{eff}\mu_B B_h V_F}{ a \hbar}  \frac{2 \sin \left ( \frac{2\pi (x-x_0)}{a}
\right
) }{\sqrt{1+\gamma ^2 + 2 \gamma \cos { \left ( \frac{2\pi (x-x_0)}{a} \right ) } }},
\end{equation}
where $V_F$ is the Fermi velocity. \\
According to Eq.~(\ref{eq:PLZ}), the probability of the Landau-Zener transitions 
$P \rightarrow 1$ if the subbands $E^{\pm}(x)$ crossover, i.e., the energy separation
$\varepsilon _{12} \rightarrow 0$. In Fig.~\ref{fig2}, we present $\varepsilon _{12}$ as a function
of the position in the nanowire and wave vector $k_x$ for two different values of $\alpha$.
We observe that for electrons with positive wave vector, i.e., flowing from the left to
the right, $\varepsilon _{12}$ reaches zero for the well-defined $k_x$,
exactly in the middle of the waveguide, i.e. for $x=0.5~\mu$m, for which $B_h^x=0$. The value of
$k_x$, for which $\varepsilon _{12}=0$, increases with the decreasing Rashba spin-orbit
coupling $\alpha$. Based on these results, we conclude that for the specified Fermi wave vector
(Fermi energy), we can tune the Rashba coupling $\alpha$ in order to reach the
Landau-Zener transition probability $P=1$. Probability $P$ as a function
of $\alpha$ is depicted in Fig.~\ref{fig3} for two different Fermi wave vectors. Fig.~\ref{fig3}
shows that the maximum of the Landau-Zener transition probability shifts towards the
higher value of $\alpha$ with decreasing Fermi wave vector $k_F$.

The analysis of the Landau-Zener transition probability presented above
is needed to understand the spin-transistor action in the proposed spin-FET. In order to
demonstrate this operation in a quantitative manner, we have performed the numerical calculations of
the conductance using the scattering matrix method 
on the square lattice with $\Delta x=\Delta y=2$~nm.\cite{kwant} We have considered the 
electron transport for a fixed Fermi energy assuming a small bias between the leads and calculated
the conductance $G=\frac{e^2}{h} \sum _i \sum _k T_{i \rightarrow k}$, where the 
summation runs over the transmission probabilities $T_{i \rightarrow k}$ from the $i$ subband on the
input to the $k$ subband in the output channel.

In the calculations, the following values of the parameters have been used: the conduction-band
electron  mass in InSb, $m^*=0.014 m_e$ where $m_e$ is the free electron mass, and $g_{eff}=-51$.
We adopt the hard-wall boundary conditions in the $y$ direction assuming the width of the conduction
channel $W=40$~nm, while the value of the helical magnetic field amplitude $B_h$ has been taken on
the basis of the experimental report~\cite{Betthausen2012} and is equal to $B_h=50$~mT.

\begin{figure}[ht]
\begin{center}
\includegraphics[scale=0.3, angle=0]{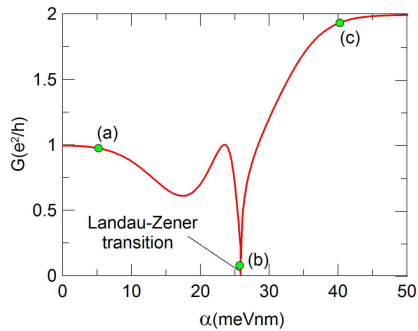}
\caption{(Color online) Conductance $G$ as a function of Rashba spin-orbit coupling
$\alpha$ calculated for Fermi energy $E_F=4.2$~meV. Points (a),(b) and (c) are chosen for the
further analysis.
}
\label{fig4}
\end{center}
\end{figure}

\begin{figure}[ht]
\begin{center}
\includegraphics[scale=0.4, angle=0]{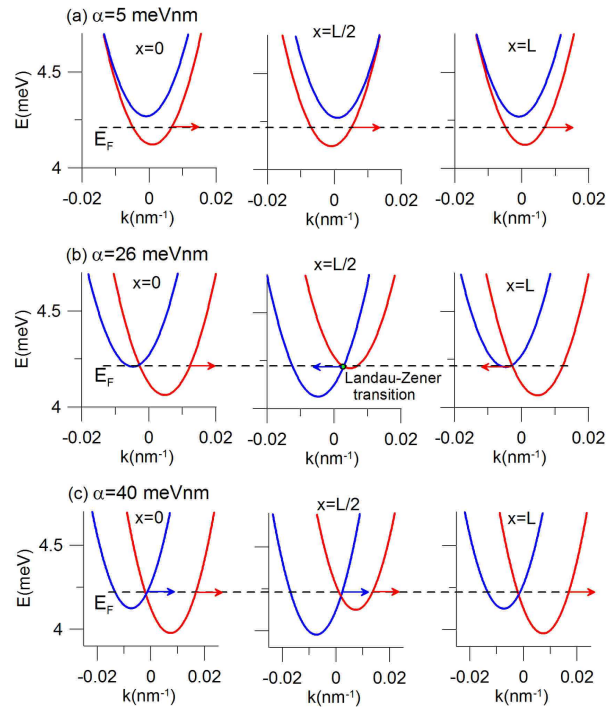}
\caption{(Color online) Conduction subbands in the left lead ($x=0$), in the  middle of the
waveguide ($x=L/2$), where the ferromagnetic stripe is located, and in the right lead ($x=L$) for
(a) $\alpha=5$~meVnm, (b) $\alpha=26$~meVnm and (c) $\alpha=40$~meVnm. These values of $\alpha$,
are marked by points in Fig.~\ref{fig4}. Fermi energy $E_F$ is marked by the dashed horizontal line.
The red (blue) curve corresponds to the dispersion relation for subband $E^+ (E^-)$. The arrows
show the direction of the electron motion.
}
\label{fig5}
\end{center}
\end{figure}

Fig.~\ref{fig4} presents the conductance as a function of the Rashba spin-orbit coupling
$\alpha$ calculated for the Fermi energy $E_F=4.2$~meV. 
As we can see, in the presence of the spin-orbit interaction the conductance exhibits the sharp
distinct dip at certain value of the spin-orbit coupling $\alpha$ [point (b)], i.e. the
device is switched into the high-resistance state. Since the changes of $\alpha$ can be realized by
the voltages applied to the side electrodes, this allows us for the full electrical control of the
transistor state, i.e. the device can be switched between the high resistance (off state) and the
low resistance (on state) states. 
The appearance of the conductance dip can be explained based on the analysis of the
electron conduction subbands, which vary with position in the waveguide. The dispersion relations 
$E^{\pm}(k)$ at the point $x$ are determined by combining the Rashba effective magnetic field
$\mathbf{B}_R$ generated by the lateral electric field $\mathbf{F}$ and directed along the $z$-axis
in the entire nanostructure, and the helical magnetic field, which varies with the position. While
the helical magnetic field generates the Zeeman spin splitting of the conduction subbands, the
spin-orbit interaction shifts the dispersion relations $E^{\pm}(k)$ in the wave vector space, in a
manner that depends on the relative alignment of $\mathbf{B}_R$ and $\mathbf{B}_h$.
In Fig.~\ref{fig5}, we present the lowest-energy subbands in the left lead ($x=0$), in
the middle of the waveguide ($x=L/2$), where the ferromagnetic stripe is located, and in the right
lead ($x=L$), for different values of the spin-orbit strength $\alpha$ marked by the points (a-c) in
Fig.~\ref{fig4}. 

For $\alpha=5$~meVnm the electrons with Fermi energy are injected into the conduction
channel from the lowest-energy subband [see Fig.~\ref{fig5}(a)]. In this case, the Rashba spin-orbit
coupling $\alpha$ is so small that the helical magnetic
field $B_h$ is much stronger than the Rashba field $B_R$. Since the Zeeman spin-splitting is only
slightly affected  by the spin-orbit interaction, the electrons are transmitted through the
nanostructures without the scatterings giving raise to the conductance $G\approx e^2/h$ (cf.
Fig.~\ref{fig4}). For $\alpha=26$~meVnm [Fig.~\ref{fig5}(b)] a degeneracy point emerges in the
middle of the waveguide, i.e., for $x=L/2$. At this point, for the chosen Fermi energy, the Rashba
field $B_R$ compensates the helical field $B_h$ and the total magnetic field vanishes. Since the
distance between spin-split eigenenergies approaches zero, the Landau-Zener transmission probability
$P \rightarrow 1$.  As a result the electrons are transmitted to the upper Zeeman subband [blue
curve in Fig.~\ref{fig5} (b)]. However, the energy of this subband, for positive wave vector in the
right lead is above the Fermi energy [cf. Fig.~\ref{fig5}(b), $x=L$], which means that there are no
available electronic states in the right lead. This leads to the backscattering of electrons.
As a result, the transport is blocked, which gives raise to the conductance dip presented
in Fig.~\ref{fig4}. The further increase of the Rashba coupling constant $\alpha$ causes that the
energy minima of both spin-split subbands are below Fermi energy and both the subbands conduct
electrons. The
conductance increases up to $G=2 e^2/h$ [cf. Fig.~\ref{fig4}, point (c)].
\begin{figure}[ht]
\begin{center}
\includegraphics[scale=0.3, angle=0]{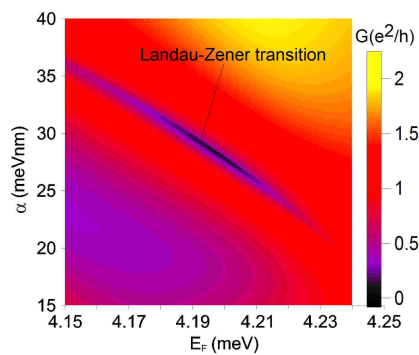}
\caption{(Color online) Conductance $G$ as a function of Rashba spin-orbit coupling
$\alpha$ and Fermi energy $E_F$.
}
\label{fig6}
\end{center}
\end{figure}

The spin-transistor action in the proposed device is generated by the Landau-Zener
transitions between spin-split subbands, which occurs when the Rashba effective field compensates
the helical field at some region of the nanostructure. For the fixed Fermi energy this can
be realized by applying the appropriate voltages to the side gate electrodes. Nevertheless, the
spin-orbit interaction strength depends not only on the electric field but also on the
Fermi energy. This means that for different Fermi energies the conditions for the Landau-Zener
transitions are fulfilled for different Rashba coupling constant. This dependence is depicted
in Fig.~\ref{fig6}, which presents the conductance as a function of coupling $\alpha$ and
Fermi energy $E_F$. In Fig.~\ref{fig6}, we mark the range, in which the Landau-Zener transitions
occur. We see that the values of $\alpha$, for which the spin-transistor action is expected,
decreases with the increasing Fermi energy $E_F$.

In summary, we have proposed the electrically controlled spin transistor setup, which is based on
the effect of the helical magnetic field generated by the ferromagnetic stripes and the lateral
Rashba spin-orbit interaction with the strength tuned by the voltages applied to the
side electrodes. 
We have shown the the appropriate tunning of the Rashba spin-orbit interaction results in
the blocking of the electron transport due to the Landau-Zener transition. This allows us for the
all-electric control of the device and switching it between the low-resistance (on state)
to the high resistance state (off state). It is worth noting that the proposed setup combines the
blocks already tested in the recent experiments: the helical magnetic field in 2DEG has been
generated by the ferromagnetic stripes used in Ref.~\onlinecite{Betthausen2012}, while the lateral
Rashba spin-orbit interaction in the QPC has been intensively studied as potential spin
filters.\cite{Debray2009,Das2011} Therefore, we expect that the proposed spin transistor can be
realized in the near future.

%

\end{document}